\documentclass[conference]{IEEEtran}
\IEEEoverridecommandlockouts

\usepackage{cite}
\usepackage{amsmath,amssymb,amsfonts}
\usepackage{algorithmic}
\usepackage{graphicx}
\usepackage{textcomp}
\usepackage[most]{tcolorbox}
\usepackage{fontawesome5}
\usepackage{lmodern}
\usepackage{array}
\usepackage{tikz}
\usepackage{adjustbox}
\usepackage{enumitem}  

\usepackage{hyperref}

\usepackage{tcolorbox}  
\usepackage{lipsum}
\usepackage{tabularx}

\tcbuselibrary{listingsutf8}

\definecolor{sectionbg}{HTML}{E9F5FF}
\definecolor{subsectionbg}{HTML}{FFFFFF}
\definecolor{borderblue}{HTML}{0078D4}
\definecolor{textblue}{HTML}{003C78}
\definecolor{refserver}{HTML}{D5E8D4}
\definecolor{thirdparty}{HTML}{FFF2CC}
\definecolor{community}{HTML}{F8CECC}
\usepackage{xcolor}

\definecolor{headcolor}{HTML}{2F4858}
\definecolor{colbg}{HTML}{F4F6F9}
\definecolor{catcolor}{HTML}{F5A623}
\renewcommand{\arraystretch}{1.3}
\newcolumntype{C}[1]{>{\centering\arraybackslash}m{#1}}

\usepackage[edges]{forest}
\tikzset{%
    parent/.style =          {align=center,text width=3.5cm,rounded corners=3pt},
    child/.style =           {align=center,text width=3.5cm,rounded corners=3pt},
    grandchild/.style =      {align=center,text width=3.5cm,rounded corners=3pt},
    greatgrandchild/.style = {align=center,text width=1.5cm,rounded corners=3pt},
    referenceblock/.style =  {align=center,text width=1.5cm,rounded corners=2pt}
}

\usepackage{comment}

\def\BibTeX{{\rm B\kern-.05em{\sc i\kern-.025em b}\kern-.08em
    T\kern-.1667em\lower.7ex\hbox{E}\kern-.125emX}}

\begin{document}


\title{
Pediatric Asthma Detection with Google’s HeAR Model: An AI-Driven Respiratory Sound Classifier
}

\author{
\IEEEauthorblockN{ Abul Ehtesham\textsuperscript{1}, Saket Kumar\textsuperscript{2}, Aditi Singh\textsuperscript{3}, Tala Talaei Khoei\textsuperscript{4}}
\IEEEauthorblockA{
\textsuperscript{1}\textit{Kent State University, USA} \\
\textsuperscript{2}\textit{Northeastern University, USA} \\
\textsuperscript{3}\textit{Department of Computer Science, Cleveland State University, USA} \\
\textsuperscript{4}\textit{Khoury College of Computer Science, Roux Institute at Northeastern University, USA} \\
 aehtesha@kent.edu, kumar.sak@northeastern.edu,\\a.singh22@csuohio.edu, 
 t.talaeikhoei@northeastern.edu
}
}
\maketitle

\begin{abstract}
Early detection of asthma in children is crucial to prevent long-term respiratory complications and reduce emergency interventions. This work presents an AI-powered diagnostic pipeline that leverages Google’s Health Acoustic Representations (HeAR) model to detect early signs of asthma from pediatric respiratory sounds. The SPRSound dataset, the first open-access collection of annotated respiratory sounds in children aged 1 month to 18 years, is used to extract 2-second audio segments labeled as wheeze, crackle, rhonchi, stridor, or normal. Each segment is embedded into a 512-dimensional representation using HeAR, a foundation model pretrained on 300 million health-related audio clips, including 100 million cough sounds. Multiple classifiers, including SVM, Random Forest, and MLP, are trained on these embeddings to distinguish between asthma-indicative and normal sounds. The system achieves over 91\% accuracy, with strong performance on precision-recall metrics for positive cases.
In addition to classification, learned embeddings are visualized using PCA, misclassifications are analyzed through waveform playback, and ROC and confusion matrix insights are provided. This method demonstrates that short, low-resource pediatric recordings, when powered by foundation audio models, can enable fast, noninvasive asthma screening. The approach is especially promising for digital diagnostics in remote or underserved healthcare settings.

\end{abstract}


\section{Introduction}

Asthma is among the most prevalent chronic respiratory diseases worldwide, affecting over 339 million individuals, with a significant number of cases occurring in individuals under 18 years of age \cite{aafa2021, Asher2021GlobalAsthma, Schluger2014Lung}. In pediatric populations, asthma is often underdiagnosed due to non-specific symptoms, variability in clinical presentation, and limited access to specialized care, especially in low-resource and rural settings. Delayed or missed diagnosis can significantly hinder timely treatment; therefore, early detection of asthma in children is crucial to prevent long-term respiratory complications and reduce emergency interventions.

This work presents an AI-powered diagnostic pipeline that leverages Google’s Health Acoustic Representations (HeAR) model \cite{baur2024hearhealthacoustic, googleHeAR2024} to detect early signs of asthma from pediatric respiratory sounds. The SPRSound dataset \cite{SPRSound}, the first open-access collection of annotated respiratory sounds in children aged 1 month to 18 years, is used to extract 2-second audio segments labeled as wheeze, crackle, rhonchi, stridor, or normal.

Each segment is embedded into a 512-dimensional representation using HeAR, a foundation model pretrained on 300 million health-related audio clips, including 100 million cough sounds. Multiple classifiers, including SVM, Random Forest, and MLP, are trained on these embeddings to distinguish between asthma-indicative and normal sounds. The system achieves over 91\% accuracy, with strong performance on precision-recall metrics for positive cases.

In addition to classification, the learned embeddings are visualized using PCA, misclassifications are analyzed through waveform playback, and ROC as well as confusion matrix analyses are provided. This method demonstrates that short, low-resource pediatric recordings, when powered by foundation audio models, can enable fast, noninvasive asthma screening. The approach is especially promising for digital diagnostics in remote or underserved healthcare settings. If left undiagnosed in early childhood, asthma can lead to long-term respiratory complications, increased hospitalizations, and reduced quality of life.

Traditional diagnostic methods for asthma, such as spirometry \cite{Roychowdhury2020} or physician-based auscultation \cite{Sarkar2015}, are either invasive, inconsistent due to clinician subjectivity, or difficult to perform reliably on young children. In this context, there is a growing interest in non-invasive, scalable, and objective approaches to respiratory health monitoring using bioacoustic signals.

Health Acoustic Representations (HeAR)  recently introduced by Google Research represent a breakthrough in bioacoustic AI modeling. Trained on over 300 million de-identified audio samples, including 100 million cough sounds, HeAR is a foundation model that learns generalized audio embeddings \cite{kohn-2015-whats} capable of capturing subtle yet meaningful health-related patterns in sound. Its ability to generalize across devices and populations makes it an ideal backbone for clinical applications in noisy real-world environments.

This paper presents a novel, AI-powered pipeline for early childhood asthma detection using Google’s HeAR model, applied to SPRSound \cite{SPRSound}, a publicly available pediatric respiratory sound dataset. The dataset comprises over 9,000 expertly annotated respiratory sound events---including wheeze, crackles, rhonchi, and stridor---recorded from children aged 1 month to 18 years. To capture relevant acoustic biomarkers, 2-second audio clips are extracted and transformed into embeddings using the HeAR model. These embeddings serve as input features for a suite of lightweight classifiers designed to identify asthma-related patterns, particularly wheeze or wheeze-like acoustic signatures.

Our approach demonstrates how foundation audio models like HeAR can be adapted for domain-specific applications, even in data-constrained healthcare settings. This research contributes a reproducible pipeline, evaluation benchmarks, and interpretability tools (e.g., waveform and spectrogram visualization of misclassifications) that together pave the way toward accessible, equipment-free asthma screening tools for children.

\section{Related Work}

\subsection{Traditional Asthma Diagnosis}
The diagnosis of asthma in children traditionally relies on clinical history, auscultation, and pulmonary function tests such as spirometry \cite{Saglani2019,MayoClinicAsthmaDiagnosis,nousias2022patientspecificmodellingsimulationrealtime}. However, spirometry \cite{lamb2023spirometry} is often impractical for children under five years old due to the need for patient cooperation. Moreover, the interpretation of lung sounds via stethoscope remains highly subjective and varies significantly across clinicians, leading to inconsistencies in diagnosis and treatment decisions. These limitations have motivated the development of automated, objective, and non-invasive diagnostic methods.

\subsection{Respiratory Sound Analysis in AI}
Over the past decade, respiratory sound analysis has gained traction as a means to automate disease detection \cite{Badnjevic2018} using machine learning (ML) \cite{fi15100332} and deep learning (DL) techniques. Approaches leveraging Mel-frequency cepstral coefficients (MFCCs) \cite{melfrequency}, wavelet transforms \cite{waveletTransform}, or spectrograms \cite{Rietveld1999} have been proposed for classifying respiratory sounds \cite{DiagnosingBreathing}, including wheeze, crackles, and rhonchi. Databases such as ICBHI 2017 \cite{HarvardDataverseDataset} and HF Lung V1 \cite{che2021hf_lung_v1} have enabled research in this area, but most focus on adult populations and lack high-quality pediatric annotations.

The SPRSound dataset, introduced by SJTU in 2022, fills this gap by offering the first open-source, pediatric respiratory sound database. It includes over 9,000 manually annotated events labeled by expert physicians and collected using digital stethoscopes in clinical environments. However, prior ML-based research on this dataset has predominantly relied on traditional classifiers trained on handcrafted features, which limits generalization and robustness in real-world scenarios.

\subsection{Foundation Models in Health Audio}
Recent advances in audio-based foundation models have transformed the processing and interpretation of acoustic signals. Notably, Google's Health Acoustic Representations (HeAR) model has emerged as a highly capable backbone for a variety of medical audio tasks. Trained on over 300 million diverse and de-identified samples, HeAR generates robust 512-dimensional embeddings from 2-second audio segments and generalizes well across acoustic domains, languages, and recording devices.

Early applications of HeAR have demonstrated its effectiveness in tasks such as cough classification, tuberculosis screening, and lung health assessment in collaboration with companies like Salcit Technologies. However, to the best of our knowledge, HeAR has not yet been applied to early asthma detection in children a critical and underexplored area in digital respiratory diagnostics.

\section{Methodology}
This section outlines the step-by-step pipeline used to develop an AI-powered classifier for early childhood asthma detection using the SPRSound dataset and Google’s HeAR model.

\subsection{Dataset: SPRSound Pediatric Respiratory Database}

The SPRSound dataset~\cite{SPRSound}, the first open-access respiratory sound database specifically focused on children aged from 1 month to 18 years, is utilized. The data were collected at the Pediatric Respiratory Department of Shanghai Children’s Medical Center (SCMC) using Yunting Model II digital stethoscopes in a clinical environment.

The dataset includes both record-level and event-level annotations, providing high-quality labels for various types of adventitious respiratory sounds. Each \texttt{.wav} file is accompanied by a \texttt{.json} file with corresponding annotations.

\begin{itemize}
    \item \textbf{2,683} respiratory recordings in \texttt{.wav} format
    \item \textbf{9,089} annotated respiratory events labeled by 11 pediatric physicians
    \item Event-level annotations include: \textit{Normal}, \textit{Wheeze}, \textit{Wheeze+Crackle}, \textit{Stridor}, \textit{Rhonchi}, \textit{Fine Crackle}, and \textit{Coarse Crackle}
    \item Record-level labels include: \textit{Normal}, \textit{Clinical Asthma Score} (CAS)~\cite{kenyon2015definition}, \textit{DAS}, \textit{CAS \& DAS}, and \textit{Poor Quality}
    \item Poor-quality recordings were flagged and excluded from this study
\end{itemize}

Each filename encodes five attributes: patient number, age, gender, recording location (e.g., posterior or lateral chest), and recording ID. The precise annotations and structured file format make this dataset an excellent benchmark for training machine learning models for pediatric respiratory diagnosis.

\subsection{Audio Segmentation and Labeling}
Each respiratory recording is paired with a JSON file containing the start and end times of annotated sound events. The following strategy is adopted to extract event-level segments:

\begin{itemize}
    \item Events shorter than 2 seconds are zero-padded to match the model’s input requirement.
    \item Longer events are sliced into overlapping 2-second clips using a sliding window.
    \item \textbf{Positive class:} \textit{Wheeze}, \textit{Wheeze+Crackle}, \textit{Rhonchi}, \textit{Stridor}~\cite{nousias2022patientspecificmodellingsimulationrealtime}
    \item \textbf{Negative class:} \textit{Normal}
\end{itemize}

All processed audio clips are saved in a new folder structure (\texttt{asthma\_clips}) and referenced in a metadata CSV file for traceability.

\subsection{Embedding Generation with Google’s HeAR Model}
To extract informative features from each audio clip, Health Acoustic Representations (HeAR), released by Google Research, is a bioacoustic foundation model trained on over 300 million audio samples and optimized for health-related sound classification.

\begin{itemize}
    \item Input: 2-second, 16kHz audio clips
    \item Output: 512-dimensional embeddings capturing temporal and spectral properties
\end{itemize}

The model is accessed from the Hugging Face Hub \cite{huggingface2025} and inference is run using a TensorFlow Serving layer. The extracted embeddings are saved in the following format:

\begin{itemize}
    \item \texttt{embeddings.npy} -- Matrix of audio embeddings
    \item \texttt{labels.npy} -- Corresponding binary labels (0: Normal, 1: Abnormal)
    \item \texttt{filenames.txt} -- Mapping of filenames for interpretability
\end{itemize}

\subsection{Classifier Training and Evaluation}
Five classical machine learning classifiers are experimented with using the extracted embeddings as input:

\begin{table}[h]
\centering
\caption{Models Used for Binary Classification}
\begin{tabular}{|l|l|}
\hline
\textbf{Model} & \textbf{Description} \\
\hline
SVM (Linear) & Support Vector Machine with linear kernel \\
Logistic Regression & Probabilistic linear classifier \\
Random Forest & Ensemble of decision trees \\
Gradient Boosting & Boosted tree-based ensemble \\
MLP Classifier & Multi-layer perceptron neural network \\
\hline
\end{tabular}
\end{table}

The dataset is split using an 80/20 stratified train-test split. For each model, Accuracy, Precision, Recall, F1-score (for both positive and negative classes), Confusion Matrix, ROC curve, and AUC are reported. All outputs, including predictions, evaluation metrics, ROC plots, and confusion matrices, are saved in the \texttt{results/} directory.

\subsection{Misclassification Analysis and Expert Review}
To support downstream validation by clinicians:

\begin{itemize}
    \item A CSV of misclassified examples with confidence scores is saved.
    \item A script allows playback of each misclassified clip using \texttt{pygame}.
    \item This facilitates human-in-the-loop review of false positives and false negatives.
\end{itemize}

Such interpretability features help build trust in AI models for medical deployment and open pathways for collaborative refinement.

\begin{figure*}[h]
    \centering
    \includegraphics[width=0.7\linewidth]{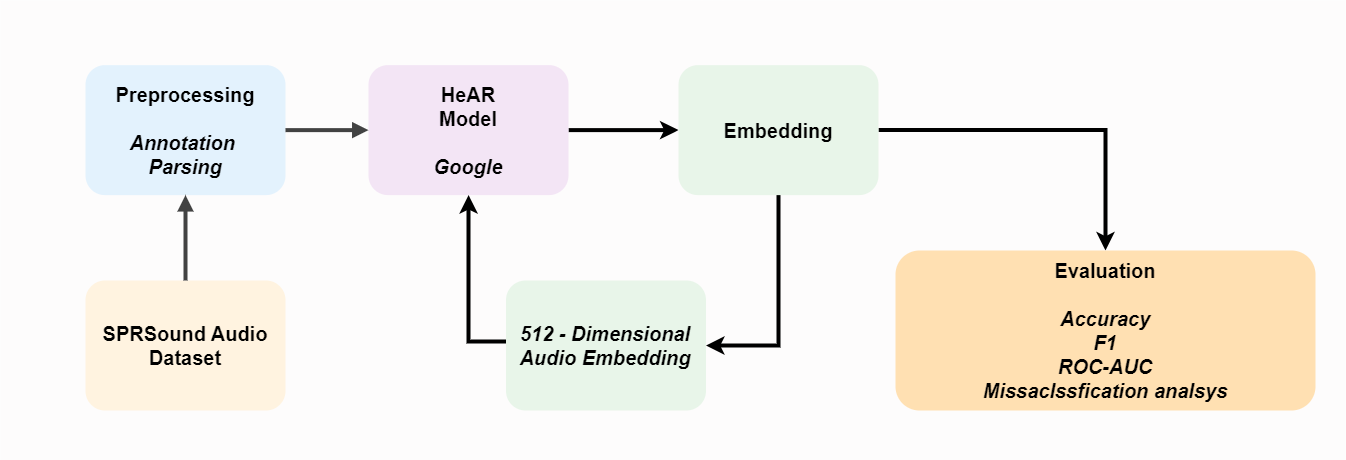}
    \caption{System architecture for AI-based asthma detection using Google’s HeAR model and the SPRSound pediatric respiratory dataset.}
    \label{fig:system_architecture}
\end{figure*}
\section{System Architecture}

Figure~\ref{fig:system_architecture} presents an overview of the end-to-end pipeline for early detection of childhood asthma using the Google HeAR model and pediatric respiratory sound data from the SPRSound dataset. The architecture is designed to support robust preprocessing, embedding generation, machine learning classification, and error analysis. Each component is modular, enabling future extension to additional disease types or deployment scenarios.

\subsection{Stage 1: Audio Preprocessing and Segmentation}
Raw respiratory recordings in WAV format are parsed using the annotation files provided in JSON format. Each recording is segmented into event-level clips (e.g., Wheeze, Normal, Rhonchi) based on start and end timestamps. Each audio segment is converted to mono, resampled to 16kHz, and standardized to a length of 2 seconds through either chunking or zero-padding. Clips labeled as “Poor Quality” are filtered out during preprocessing, ensuring that only medically valid and noise-tolerant segments are passed to the embedding model.

\subsection{Stage 2: Embedding Extraction via HeAR Model}

All preprocessed 2-second clips are fed into the Health Acoustic Representations (HeAR) model released by Google Research. This model produces a 512-dimensional embedding vector for each clip, capturing the latent acoustic features associated with respiratory anomalies such as wheezing or stridor. These embeddings are generated using the TFSMLayer interface in TensorFlow, ensuring compatibility with the Hugging Face model checkpoint for ``google/hear'' \cite{google_hearhub}. Embeddings and associated metadata (label, filename, source) are saved for downstream model training and interpretability.

\subsection{Stage 3: Supervised Classifier Training}

The embeddings are split into training and testing sets (80\%/20\%) while preserving label distribution. A suite of machine learning classifiers—including Support Vector Machine (SVM), Logistic Regression, Random Forest, Gradient Boosting, and Multi-layer Perceptron (MLP) \cite{fi15100332}—are trained using the 512-dimensional HeAR vectors as input. All models are implemented via Scikit-learn and evaluated using standard metrics: accuracy, F1-score, precision, recall, and Receiver Operating Characteristic Area Under the Curve (ROC-AUC).

\subsection{Stage 4: Evaluation and Error Analysis}

The trained classifiers are evaluated on the test set. Confusion matrices and Principal component analysis (PCA) based visualizations are generated to assess class separability. To support real-world usability, a listening interface is provided for clinicians to listen to misclassified clips (false positives and false negatives). This enables transparent evaluation and cross-validation with human interpretation, a key factor in clinical acceptance.


\section{Experiments and Evaluation}

\subsection{Experimental Setup}

To evaluate the childhood asthma detection pipeline, the following setup was utilized:

\begin{itemize}
    \item \textbf{Frameworks and Libraries:} TensorFlow (TFSMLayer) \cite{tensorflow2025}, Hugging Face Hub for model access, Librosa \cite{librosa2025} for audio preprocessing, and Scikit-learn \cite{scikit-learn2025} for machine learning and evaluation.
    \item \textbf{Audio Preprocessing:} Each respiratory sound event was converted to a mono-channel waveform and resampled to 16kHz. Audio segments were extracted in fixed windows of 2 seconds (i.e., 32,000 samples).
    \item \textbf{Dataset:} The SPRSound pediatric respiratory dataset was used. Poor-quality recordings were filtered out, and only labeled segments corresponding to \texttt{Wheeze}, \texttt{Wheeze+Crackle}, \texttt{Rhonchi}, \texttt{Stridor}, and \texttt{Normal} were retained. The final dataset consisted of 3,319 labeled clips.
    \item \textbf{Embedding Generation:} Each clip was passed through Google's HeAR model to extract a 512-dimensional audio embedding.
\end{itemize}

\subsection{Model Training and Comparison}

Five classical machine learning models were trained and compared using the extracted embeddings:

\begin{itemize}
    \item Support Vector Machine (Linear Kernel)
    \item Logistic Regression
    \item Random Forest (100 estimators)
    \item Gradient Boosting (100 estimators)
    \item Multi-Layer Perceptron (128 and 64 hidden units)
\end{itemize}

All models were trained on 80\% of the dataset, and evaluated on a held-out 20\% test set, with stratified sampling to preserve label distribution.

The following metrics were used for evaluation:

\begin{itemize}
    \item \textbf{Accuracy:} Proportion of correctly classified samples.
    \item \textbf{Precision, Recall, F1-score:} Computed for the \texttt{positive} (asthma) class.
    \item \textbf{ROC-AUC:} Area under the receiver operating characteristic curve.
\end{itemize}

\subsection{Results Summary}



\begin{figure}[ht]
    \centering
    \includegraphics[width=\linewidth]{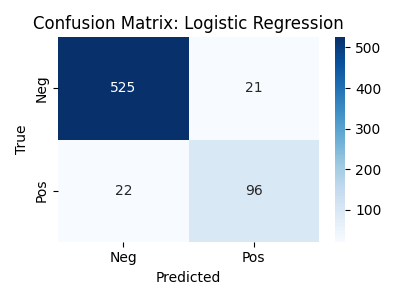}
    \caption{Confusion Matrix: Logistic Regression}
    \label{fig:cm_logistic_regression}
\end{figure}
\begin{figure}[ht]
    \centering
    \includegraphics[width=\linewidth]{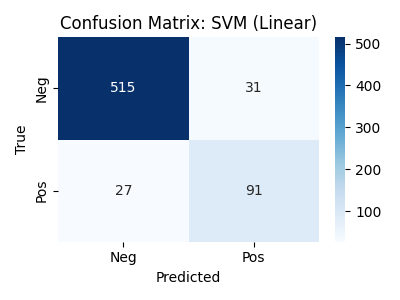}
    \caption{Confusion Matrix: SVM (Linear)}
    \label{fig:cm_svm_linear}
\end{figure}
\begin{figure}[ht]
    \centering
    \includegraphics[width=\linewidth]{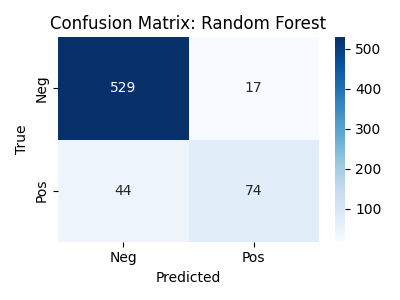}
    \caption{Confusion Matrix: Random Forest}
    \label{fig:cm_random_forest}
\end{figure}
\begin{figure}[ht]
    \centering
    \includegraphics[width=\linewidth]{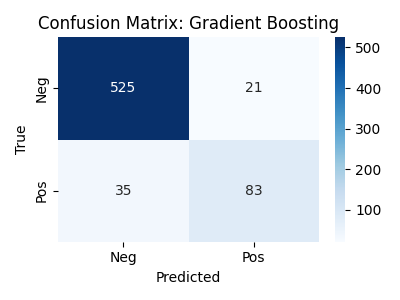}
    \caption{Confusion Matrix: Gradient Boosting}
    \label{fig:cm_gradient_boosting}
\end{figure}
\begin{figure}[ht]
    \centering
    \includegraphics[width=\linewidth]{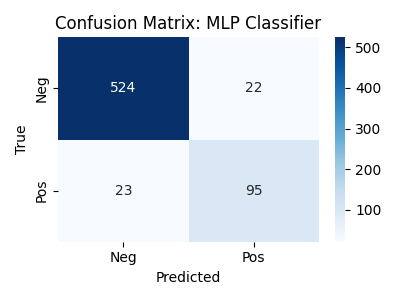}
    \caption{Confusion Matrix: MLP Classifier}
    \label{fig:cm_mlp_classifier}
\end{figure}

\vspace{1em}
\noindent
\textbf{Confusion Matrices:} To evaluate model performance, confusion matrices were analyzed for each classifier. These matrices highlight false positives—healthy samples misclassified as asthmatic—and false negatives—asthmatic samples misclassified as healthy.

:

\begin{itemize}
    \item \textbf{Gradient Boosting (Figure~\ref{fig:cm_gradient_boosting}):} Moderate performance with 35 false negatives and 21 false positives.
    \item \textbf{Logistic Regression (Figure~\ref{fig:cm_logistic_regression}):} Strong performance, showing the lowest false negatives (22) among all models.
    \item \textbf{MLP Classifier (Figure~\ref{fig:cm_mlp_classifier}):} Similar to Logistic Regression, with 23 false negatives and 22 false positives.
    \item \textbf{Random Forest (Figure~\ref{fig:cm_random_forest}):} Higher false negatives (44), indicating reduced sensitivity to positive (asthmatic) cases.
    \item \textbf{SVM Linear (Figure~\ref{fig:cm_svm_linear}):} Balanced results with 27 false negatives and 31 false positives.
\end{itemize}

These matrices provide insight into each model’s classification tendencies, especially their ability to correctly identify asthmatic versus non-asthmatic samples.

\begin{figure}[ht]
    \centering
    \includegraphics[width=0.8\linewidth]{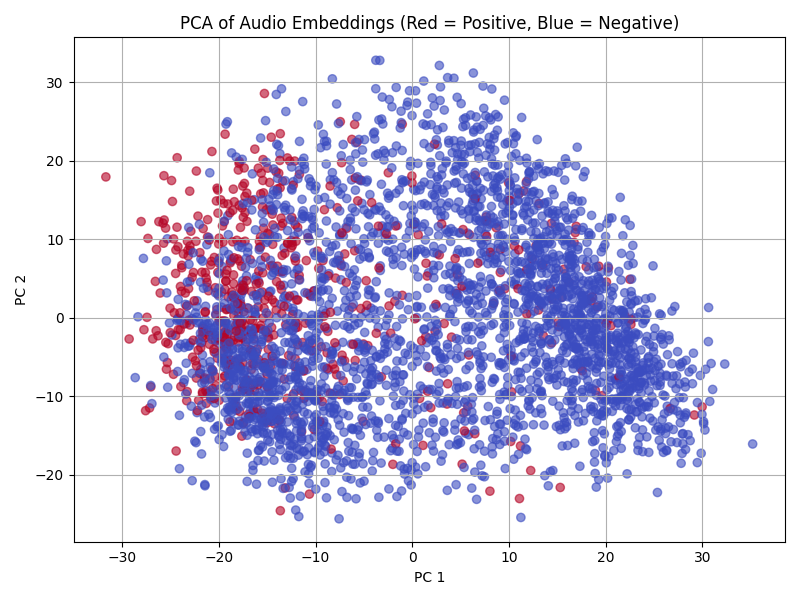}
    \caption{PCA of Audio Embeddings (Red = Positive, Blue = Negative)}
    \label{fig:pca_embeddings}
\end{figure}

\noindent
\textbf{PCA Visualization:} To visualize class separability in the embedding space, the 512-dimensional HeAR embeddings were projected to 2D using PCA. Figure~\ref{fig:pca_embeddings} shows the distribution of positive (red) and negative (blue) audio samples.

\begin{figure}[ht]
    \centering
    \includegraphics[width=0.8\linewidth]{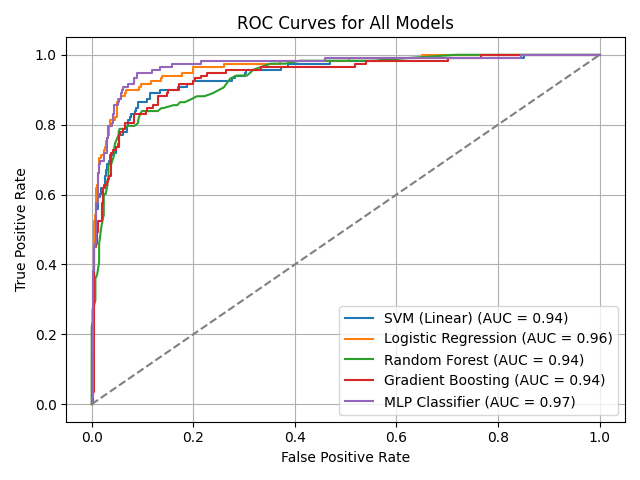}
    \caption{ROC curves for all models. The MLP Classifier outperforms others with the highest AUC score of 0.97.}
    \label{fig:roc_curves}
\end{figure}

\noindent
\textbf{ROC Curves:} To evaluate model performance in terms of classification thresholds, ROC curves were generated for all classifiers. These plots illustrate the trade-off between true positive rate and false positive rate across different thresholds. As shown in Figure~\ref{fig:roc_curves}, the MLP Classifier achieved the highest AUC (0.97), followed by Logistic Regression (0.96), while SVM (Linear), Random Forest, and Gradient Boosting each achieved an AUC of 0.94.

\begin{table}[ht]
\centering
\caption{Performance of Classifiers on Childhood Asthma Detection}
\renewcommand{\arraystretch}{1}     
\setlength{\tabcolsep}{3pt}            
\footnotesize                          
\begin{tabular}{lcccc}
\hline
\textbf{Model} & \textbf{Accuracy} & \textbf{Precision$_{\text{pos}}$} & \textbf{Recall$_{\text{pos}}$} & \textbf{F1$_{\text{pos}}$} \\
\hline
SVM (Linear)        & 0.91 & 0.75 & 0.77 & 0.76 \\
Logistic Regression & 0.94 & 0.82 & 0.81 & 0.81 \\
Random Forest       & 0.91 & 0.79 & 0.66 & 0.72 \\
Gradient Boosting   & 0.91 & 0.78 & 0.70 & 0.74 \\
MLP Classifier      & 0.93 & 0.85 & 0.91 & 0.88 \\
\hline
\end{tabular}
\label{tab:results_summary}
\end{table}

\subsection{Error Analysis}

To better understand model limitations and real-world applicability:

\begin{itemize}
    \item A CSV of misclassified audio clips was saved for clinical review.
    \item An interactive playback interface was developed that allows clinicians to listen to misclassified examples.
    \item Some false positives corresponded to Rhonchi or Stridor, which may acoustically resemble wheezing, suggesting a need for finer-grained multi-class labeling.
\end{itemize}

\subsection*{E. Embedding Visualization and Interpretation}

\begin{figure}[!t]
\centerline{\includegraphics[width=\linewidth]{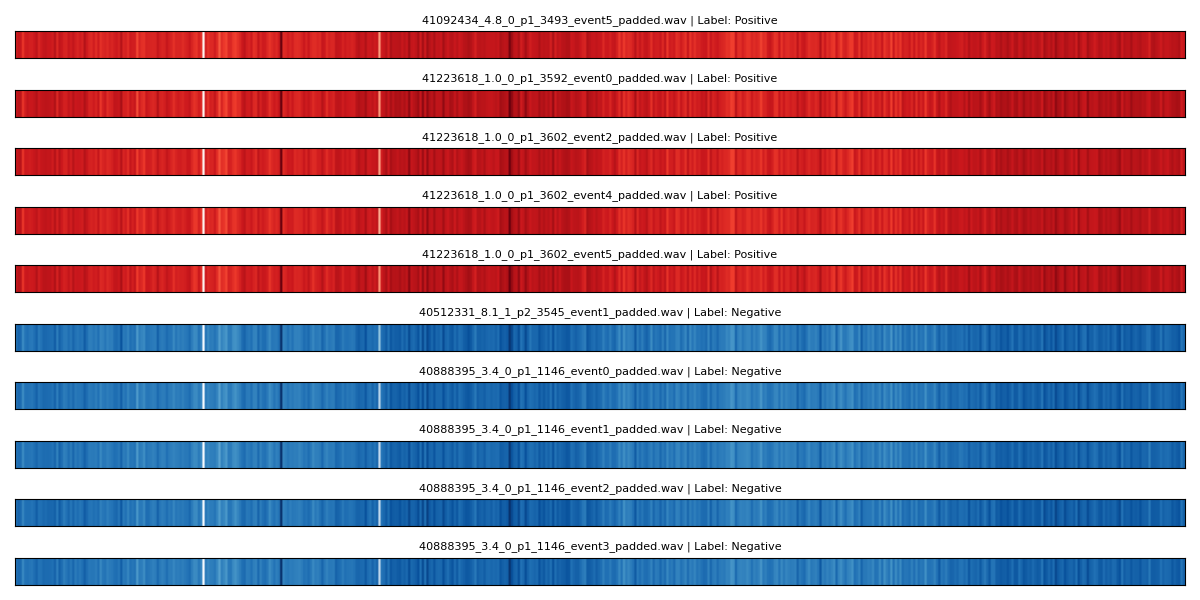}}
\caption{Barcode visualization of HEAR embeddings for selected respiratory clips. Each row corresponds to a 512-dimensional embedding. Red rows represent positive (asthmatic) clips, and blue rows represent negative (normal) clips. Note the visually distinguishable intensity patterns between categories, which highlight how HeAR captures relevant acoustic features.}
\label{fig:barcode}
\end{figure}

To interpret the embeddings, a barcode-style heatmap was used to visualize positive (red) and negative (blue) samples (Figure~\ref{fig:barcode}). Each row represents a 512-dimensional embedding, with color intensity indicating feature magnitude. Distinct activation patterns in positive samples suggest that HeAR captures disease-relevant acoustic features, providing interpretability alongside quantitative performance.

\section{Discussion}

This study presents a scalable and efficient approach to early asthma detection in children using respiratory sound analysis and Google’s Health Acoustic Representations (HeAR) model. Through rigorous experimentation on the pediatric SPRSound dataset, it was demonstrated that even short 2-second audio segments carry sufficient bioacoustic information to train highly accurate classifiers.

\subsection{Interpretation of Results}
The experiments yielded strong classification performance across all tested models, with Logistic Regression and MLP achieving the best balance between precision and recall. Confusion matrices and PCA visualizations confirmed that embeddings from HeAR provide strong separability between normal and wheezing sounds.

\subsection{Significance of HeAR}
The HeAR model, pre-trained on over 300 million audio segments by Google, offers robust generalization across microphones and respiratory conditions. This eliminates the need for large-scale domain-specific training and makes it ideal for healthcare contexts with limited annotated data.

\subsection{Benefits of Short Clip-Based Diagnosis}
Segmenting long respiratory recordings into overlapping 2-second clips improves data granularity, increases sample diversity, and enables fine-grained event detection. The pipeline capitalizes on this by filtering out poor-quality or silent segments, thus ensuring consistent embedding quality.

\subsection{Deployment Potential}
The system is lightweight and can run on local devices or cloud environments. Given the growing ubiquity of smartphones and digital stethoscopes, this framework can be embedded into telehealth solutions or integrated into low-resource clinics for community-based screenings.

\subsection{Limitations}
Despite strong performance, false positives and false negatives still occur. Some misclassifications arose from ambiguous sound types like Rhonchi, which can overlap acoustically with wheezing. Furthermore, background noise, especially in real-world environments, could affect embedding quality.

\subsection{Clinical Validation}
A listening interface for medical professionals to review misclassified clips was incorporated. This tool enables subjective verification, fosters transparency, and supports human-in-the-loop validation for eventual clinical integration.

\subsection{Future Work}
Beyond asthma, the SPRSound dataset also includes annotations for other adventitious sounds such as Stridor, Crackles, and Rhonchi. Future directions include adapting this pipeline to detect pneumonia or TB, validating cross-device robustness, and enhancing performance through multi-label and semi-supervised learning techniques.

\section{Conclusion}

This study presents an AI-driven pipeline for early detection of childhood asthma using pediatric respiratory audio.  Leveraging the Google HeAR model for acoustic embedding generation and applying classical machine learning classifiers, the system demonstrates strong performance in distinguishing between normal and wheezing sounds. Evaluation on the SPRSound dataset, supported by PCA visualization and confusion matrix analysis, confirms the effectiveness of the proposed approach. Future work will focus on expanding to multi-disease detection, analyzing temporal trends, and improving cross-device generalization to enhance real-world clinical applicability.

\bibliographystyle{IEEEtran}
\bibliography{refs}


\end{document}